\documentclass[authoryear,preprint,12pt,authoryear]{elsarticle}

\usepackage{amssymb,amsmath}
\usepackage[latin1]{inputenc}   

\usepackage{subfigure}

\newcommand{\A}{\boldsymbol{A}}
\newcommand{\B}{\boldsymbol{B}}
\newcommand{\ALPHA}{\boldsymbol{\alpha}}
\newcommand{\BETA}{\boldsymbol{\beta}}
\newcommand{\e}{\boldsymbol{e}}
\newcommand{\EPS}{\boldsymbol{\varepsilon}}
\newcommand{\g}{\boldsymbol{\gamma}}
\newcommand{\G}{\boldsymbol{\Gamma}}
\newcommand{\I}{\boldsymbol{I}}
\newcommand{\M}{\boldsymbol{M}}
\newcommand{\MU}{\boldsymbol{\mu}}
\newcommand{\PI}{\boldsymbol{\Pi}}
\newcommand{\x}{\boldsymbol{x}}
\newcommand{\y}{\boldsymbol{y}}
\newcommand{\z}{\boldsymbol{z}}

\journal{Energy Economics}

\begin{document}

\begin{frontmatter}



\title{Causal modeling and inference for electricity markets}


\author{Egil Ferkingstad\corref{cor1}}\ead{egil.ferkingstad@nr.no}
\author{Anders L{\o}land}\ead{anders.loland@nr.no}
\author{Mathilde Wilhelmsen}\ead{mathilde.wilhelmsen@nr.no}

\cortext[cor1]{Corresponding author. Tel.: +47 22852500; fax: +47 22697660.}

\address{Norwegian Computing Center, Post Office Box 114 Blindern, NO-0314 Oslo, Norway}

\begin{abstract}

How does dynamic price information flow among Northern European electricity  spot prices and prices of major electricity  generation fuel sources? We use time
series models combined with new advances in causal
inference to answer these questions. Applying our methods to weekly 
Nordic and German electricity prices, and oil, gas and coal prices, with
 German wind power and Nordic water reservoir levels as exogenous variables, we estimate a causal
model for the price dynamics, both for contemporaneous and lagged
relationships. 
In contemporaneous time, Nordic and German electricity prices are interlinked through gas prices. In the long run, electricity prices and British gas prices adjust themselves to establish the
equlibrium price level, since oil, coal, continental gas and EUR/USD are found to be weakly
exogenous.

\end{abstract}


\end{frontmatter}


\section{Introduction}
\label{sec:intro}

There is an ongoing debate on the convergence of and price dynamics in
energy markets, and electricity markets in particular.  For the US market, \cite{park2006pda} use advances in causal
flow modelling and find that the dynamic
relationships between electricity markets not only are governed by transmission
lines, but also by different market structure and regulation. Using
similar techniques, the same authors indicate that the Canadian and US natural gas market is a single
highly integrated market
\citep{park2008pia}.
\cite{mjelde2009mia} go one step further, and investigate how
weekly dynamic price information flows among major US electricity generation
fuel sources: natural gas, uranium, coal and crude oil.  They find
that peak electricity prices move natural gas prices, which in turn
influence crude oil prices.

To our knowledge, price dynamics among Northern European
electricity markets and their major fuel sources has not been
closely looked into before.
\cite{zachmann2008ewm} rejects the hypothesis of full market
integration of Northern European electricity 
markets. We will focus on the Nordic and German electricity
markets \citep{weron2006}. The Nordic electricity market (Nord Pool)
is dominated by highly flexible hydro power (54\% in 2007 \citep{fridolfsson09}), and even though
congestion within the Nord Pool area is not uncommon
\citep{marckhoff09}, we will consider the common Nordic system spot
price here. The German EEX market, being the largest market in Europe,
 is on the other hand dominated by coal (47\%) and nuclear power (23\%) \citep{brunekreeft05}. 
 Gas (17\%), hydro and an increasing wind power production
 complement the picture. The  EEX market is generally 
assumed to be less mature than the Nordic market
\citep{weron2006,weight08,musgens06,fridolfsson09}. 

We investigate the price dynamics between electricity prices and major fuel sources (oil, gas, coal) by estimating a causal model for the price dynamics, where Nordic water reservoir levels and German electricity production from wind mills are treated as exogenous variables.  
\cite{mjelde2009mia} estimate a vector error correction model (VECM)
for logarithmic prices. A directed acyclic graph (DAG)~\citep{spirtes2000} representing instantaneous causal influences is then found from the resulting contemporaneous correlation matrix, using the greedy equivalence search (GES) algorithm of~\cite{chickering03}. 

Most causal DAG learning algorithms, including the GES algorithm, are based on the assumption that variables are jointly normally distributed. These methods share a fundamental problem: Several DAGs usually correspond to same joint distribution, so one only obtains an equivalence class of DAGs that are indistinguishable from data. While some directions of causal influences (edges in the DAG) may be the same for all DAGs in the equivalence class, usually many or most directions are left undetermined. 

In the present paper, we rely on the assumption of \emph{non-normality}, using the linear
 non-Gaussian acyclic model (LiNGAM) recently developed by \citet{shimizu2006lng,shimizu2006fco}. This allows us to identify one single DAG. Because of this, we are also able to coherently integrate both contemporaneous and time-lagged causal relationships into the same DAG analysis. For our data, the GES algorithm is only able to identify undirected contemporaneous associations. The LiNGAM approach, on the other hand, provides instantaneous and time-lagged directed causal influences.

\section{Methods}
\label{sec:methods}

The three basic building blocks of our data analysis are the
\rm{vector autoregression} (VAR) model, the \rm{vector error correction model} (VECM) and the \rm{linear
  non-Gaussian acyclic model} (LiNGAM) \citep{shimizu2006lng,shimizu2006fco}. We will now describe
each in turn, before we  combine them to estimate both
instantaneous and lagged causal effects.

\subsection{Vector autoregression model}
\label{sec:var}

The vector autoregression model \citep{HamiltonTSA} is a standard tool of
econometrics and multivariate time series analysis. Let the
\rm{endogenous variables} $\boldsymbol x_t$ and the \rm{exogenous
  variables} $\boldsymbol  z_t$ be observed
random vectors depending on (time) $t=1,2,\ldots$.  The basic idea of
the VAR model is that the endogenous variables depend linearly on
their $k$
previous values, as well as the current value of the exogenous
variables, i.e.
\begin{align}
\boldsymbol  x_t = \boldsymbol{\mu} + \sum_{\tau=1}^k \boldsymbol{M}_{\tau}
\boldsymbol x_{t-\tau} + \boldsymbol \gamma \boldsymbol z_t + \boldsymbol
e_t,
\label{eq:var}
\end{align}
where $ \boldsymbol{M}_{\tau}$ and $ \boldsymbol \gamma $ are coefficient
matrices of size $n\times n$ and $n\times d$, respectively, where
$n$ is the number of endogenous variables and $d$ is the number of
exogenous variables. Further, $\boldsymbol{\mu}$ is a constant vector
and $\boldsymbol e_t$ is a vector of residuals (innovations). 

 All variables must have the same order of integration. If all variables are stationary, $I(0)$, we have the standard case of a VAR model. If all variables are non-stationary, $I(d)$, $d>1$, there are two possibilities. First, if the variables are not cointegrated, the variables must be differenced $d$ times in order to obtain a VAR. Second, if the variables are cointegrated, we may use a vector error correction model (VECM).  


\subsection{Vector error correction model}
\label{sec:vecm}


We here consider the case where the variables $\boldsymbol x_t$ are $I(1)$, so that they are differenced one time in order to achieve stationarity. The vector error correction model (VECM) can be derived from the VAR model in \eqref{eq:var},
\begin{align}
\Delta \x_t = \MU + \PI \x_{t-1} + \sum_{\tau=1}^{k-1}\G_{\tau} \Delta \x_{t-\tau} + \g \z_t + \e_t,
\label{eq:vecm}
\end{align}
where $\Delta$ is the difference operator ($\Delta \x_t = \x_t - \x_{t-1}$), and $\G_{\tau}$ is an $n \times n$ matrix relating changes in $\x_t$ for lagged $\tau$ periods to current changes in $\x_t$.

The matrix $\PI$ is called an error correction term, which compensates
for the long-run information lost through differencing
\citep{Juselius2006}.  $\PI=\ALPHA \BETA'$, where $\ALPHA$
and $\BETA$ are of dimensions $n \times r$, where the rank $r$ is the number of cointegration relationships. The $r$ linearly independent columns of $\BETA$ are the cointegrated vectors, each representing one long-run relationship between the series, and $\BETA'\x_{t-1}$ is then stationary. 


If $r=0$, the matrix $\PI$ does not exist, and we have a VAR in difference, not a VECM. If we have full rank, $r=n$, it does not make sense to specify the model as a VECM, as the stationary $\Delta \x_t$ in \eqref{eq:vecm} will be equal to a non-stationary $\PI \x_{t-1}$ plus some lagged stationary variables and so on, which is inconsistent \citep{Juselius2006}. 



Comparing \eqref{eq:var} with \eqref{eq:vecm} gives
\begin{align}
\PI = \ALPHA \BETA' = -(\I-\sum_{\tau=1}^{k}\M_{\tau}),
\label{eq:pi}
\end{align}
and
\begin{align}
\G_{\tau} = -\sum_{i=\tau+1}^k\M_i.
\label{eq:gamma}
\end{align}

\subsection{Linear non-Gaussian acyclic causal model}
\label{sec:lingam}

In general, a linear causal model on the zero-mean (centered) random variables $y_i$, $i=1,\ldots,m$,  can be defined by
\begin{equation}
y_i=\sum_{k(i)<k(j)} \beta_{ij} y_j + \varepsilon_i,
\label{eq:lingam}
\end{equation}
where the $\varepsilon_i$s are random noise terms and $k$ is a permutation over $\{1,\ldots,m\}$. 
We interpret $k$ as a \rm{causal ordering} of the variables, where later variables cannot cause earlier variables. Equation~\eqref{eq:lingam} can be represented as a \rm{directed acyclic graph} (DAG) with vertices corresponding to $y_i$ and edges corresponding to a nonzero $\beta_{ij}$. Estimating causal DAGs from observational data has received considerable interest in recent years~\citep{pearl2000,spirtes2000}. For continuous $y_i$, standard methods assume that the noise terms $\varepsilon_i$ are jointly normally distributed, and use the estimated covariance matrix to infer the DAG. 

Several methods for inferring DAGs from Gaussian observational data have  been
proposed. To enable comparison of our results and those
of~\citet{park2008pia}, we employ the Greedy Equivalence Search
(GES) algorithm of~\citet{chickering03}, as implemented in the
software \cite{tetradManual}. The GES algorithm uses a score to
evaluate how well a suggested DAG fits the data. Starting with an
empty DAG, a greedy search over equivalence classes (defined below) is done. The
search concludes when
local maximum of the score is reached.

A general problem with the normality-based methods such as the GES algorithm is that, even with an infinite amount of data, 
one cannot identify a unique causal model (DAG), only a so-called Markov equivalence class consisting of several different DAGs corresponding 
to the same joint distribution.
This is easily seen in the case of two variables $y_1$ and $y_2$, where there is clearly no way of distinguishing between the models $y_1 \rightarrow y_2$ and $y_1 \leftarrow  y_2$ based on the covariance structure alone, and $\{y_1 \rightarrow y_2, y_1 \leftarrow  y_2\}$ is a Markov equivalence class. With three variables, the Markov equivalence classes are $\{y_1 \rightarrow y_2 \rightarrow y_3, y_1 \leftarrow y_2 \leftarrow y_3, y_1 \leftarrow y_2 \rightarrow y_3\}$ and $\{y_1 \rightarrow y_2 \leftarrow y_3\}$. For an extensive discussion of this problem, see~\cite{shimizu2006fco}.

In contrast, when assuming that the noise terms are \rm{independent} and \rm{non-Gaussian}, a unique causal structure is in fact identifiable. Equation~\eqref{eq:lingam} is then known as the LiNGAM~\citep{shimizu2006lng}. Writing \eqref{eq:lingam} in matrix form:
\begin{equation}
  \label{eq:matlingam}
  \y = \B\y + \EPS
\end{equation}
where $\y=(y_1,\ldots,y_m)'$, $\EPS=(\varepsilon_1,\ldots,\varepsilon_m)$ and $\B$ is the (permutable to lower triangular) matrix of coefficients $\beta_{ij}$. 
The independence of the elements of $\EPS$ implies that there are "no unobserved confounders" in the sense of~\citet{pearl2000}, so a causal interpretation is valid (cf.~\citet{shimizu2006lng}, Section 2).
Letting $\A=(\I-\B)^{-1}$, we can rewrite~\eqref{eq:matlingam} as 
\begin{equation}
  \label{eq:ica1}
  \y=\A\EPS.
\end{equation}
Since the variables in $\EPS$ are independent and non-Gaussian, \eqref{eq:ica1} defines the \rm{Independent Component Analysis} (ICA) model~\citep{comon94,hyvarinen2000ica}. 

In ICA, the goal is to estimate both the so-called \rm{mixing matrix} $\A$ and the  \rm{independent components} $\EPS$. Essentially, in ICA we aim to find $\A$ and  $\EPS$ such that the entries of $\EPS$ are as statistically independent as possible. 
By an argument based on the central limit theorem, this problem can also be posed as finding components which are as non-Gaussian as possible. Non-Gaussianity can be measured using the concept of \rm{entropy}. The entropy of a random vector $\y$ with density $f$ is defined as $H(\y)=-\int\!f(\y) \log f(\y) \mathrm{d}\y$. Among random variables with a given variance, Gaussian variables have the highest possible entropy. Therefore, we can measure non-Gaussianity based on \rm{negentropy} $J$, which is defined by $J(\y)=H(\y_{g})-H(\y)$, where $\y_g$ is a Gaussian random vector having the same covariance matrix as $\y$. Clearly, $J(\y)$ is zero for Gaussian $\y$ and positive for non-Gaussian $\y$. The iterative fixed-point algorithm \rm{fastICA}~\citep{hyvarinen99fica} estimates $\A$ efficiently and robustly based on approximations to negentropy. 

It can be seen from \eqref{eq:ica1} that both $\A$ and $\EPS$ can only be estimated up to a
scaling constant and a permutation.
However, both the scaling and the permutation can be found in the application of ICA to LiNGAM, as shown by~\citet{shimizu2006lng}.
After estimating $\A$, the coefficient matrix $\boldsymbol{B}$ is immediately available as $\I - \A^{-1}$.

\subsection{Combining instantaneous and lagged effects}
\label{sec:instlag}

Our interest is here in the following model:
\begin{align}
\x_t = \MU + \sum_{\tau=0}^{k}\B_{\tau}\x_{t-\tau} + \g \z_t + \EPS_t.
\label{eq:causalModel}
\end{align}
The difference between~\eqref{eq:causalModel} and the VAR model
defined in~\eqref{eq:var} is the inclusion of instantaneous causal
effects $\B_0$, where the matrix $\B_0$ corresponds to a DAG (i.e., can be
permuted to strict lower triangularity) as in
Section~\ref{sec:lingam}. $\B_1, \B_2,\ldots$  contain autoregressive effects, and their corresponding graphs may be cyclic.  

To estimate the model in \eqref{eq:causalModel}, we customise the method described by \cite{hyvarinen2008cmc}:
\begin{enumerate}
\item Estimate a VECM model for the data, see \eqref{eq:vecm}. We here obtain the coefficient matrices $\widehat{\PI}$ and $\widehat{\G}_1,\ldots,\widehat{\G}_{k-1}$, together with $\widehat{\MU}$ and $\widehat{\g}$.
\item Translate the estimated VECM coefficients into a VAR representation, see \eqref{eq:pi} and \eqref{eq:gamma}. We then obtain the coefficient matrices $\widehat{\M}_1,\ldots,\widehat{\M}_k$ in \eqref{eq:var}.
\item Compute the residuals $\widehat{\e}_t$,
\begin{align}
\widehat{\e}_t = \x_t-\widehat{\MU}-\widehat{\g}\z_t - \sum_{\tau=1}^{k}\widehat{\M}_{\tau}\x_{t-\tau}.
\end{align}
\item Perform the LiNGAM analysis on the residuals to find an estimate of the instantaneous effect matrix $\B_0$. This matrix is a solution to the model
\begin{align}
\widehat{\e}_t = \B_0\widehat{\e}_t+\tilde{\EPS}_t.
\label{eq:lingam_oppskrift}
\end{align}
See Section \ref{sec:lingam} for details.
\item Compute the matrices of lagged causal effects, $\B_{\tau}$, $\tau>0$, which are given as
\begin{align}
\widehat{\B}_{\tau} = (\I-\B_0)\widehat{\M}_{\tau}.
\label{eq:B_lagged}
\end{align}
\end{enumerate}

How do we find \eqref{eq:B_lagged}? Equation \eqref{eq:causalModel} gives
\begin{align*}
(\I-\B_0)\x_t = \MU + \sum_{\tau=1}^{k}\B_{\tau}\x_{t-\tau} + \g \z_t + \EPS_t.
\end{align*}
This gives
\begin{align}
\begin{split}
\x_t = & \ (\I-\B_0)^{-1}\MU + \sum_{\tau=1}^{k}(\I-\B_0)^{-1}\B_{\tau}\x_{t-\tau} \\
& + (\I-\B_0)^{-1}\g \z_t + (\I-\B_0)^{-1}\EPS_t.
\label{eq:Mtderive}
\end{split}
\end{align}
Comparing \eqref{eq:Mtderive} with \eqref{eq:var}, we find that $(\I-\B_0)^{-1}\B_{\tau}=\M_{\tau}$ for $\tau\ge 1$. Also, we see that $(\I-\B_0)^{-1}\EPS_t=\e_t$, which gives rise to (\ref{eq:lingam_oppskrift}).


\subsection{Time-lagged causal flow and Granger causality}
\label{sec:granger}
One may ask whether a time-lagged causal flow is different from Granger causality. 
Granger causality is the ability to reduce the prediction error \citep{HamiltonTSA}. Based on the VAR representation \eqref{eq:var}, a variable $i$ Granger causes the variable $j$  if at least one of the coefficients of 
$\M_\tau$  from $\x_{t-\tau}^{(i)}$, $\tau\ge1$, to $\x_{t-\tau}^{(j)}$ is significantly non-zero, since this reduces the prediction error in $\x_{t}^{(j)}$. \cite{hyvarinen2008cmc} proposed a combined definition of Granger causality: If at least one of the coefficients 
$\B_\tau(j,i)$, $\tau\ge 0$,  is significantly non-zero, variable $i$ causes  $j$. See \cite{hyvarinen2008cmc} and \cite{zhang09} for a more thorough discussion.

\section{Data}
\label{sec:data}

We focus on the Nordic and German electricity markets, their major
fuel sources (gas, coal, oil) and physical variables known to partly
explain Nordic and German electricity prices; German wind power
production and Nordic water reservoir levels. The data consist of 365
weekly observations of each variable from 2002--2008. Ideally, we should have used data
further back. However, firstly, the wind data were not available until
2002. Secondly, six years is a long time in quite rapidly evolving and
increasingly integrated European gas and electricity markets
\citep{zachmann2008ewm,bunn10,micola07}. The markets were less mature further
back, but if wind data had been available, we could  have
included 2001 data as well.

All price series are given in or converted to EUR.
Transforming all prices to a common currency \citep{hovanov04} could induce
  dependencies related to exchange rate fluctuations and not
  energy price fluctuations. 
For that reason, and since exchange
rates may also influence commodity prices
\citep{chen07,farooq09,zhang08}, we include the EUR/USD exchange
rate as well.

All price series are given as averages\footnote{
Using weekly average spot prices might introduce additional correlation
into the series or differenced price series \citep{working60}. 
Under some applications one might want to use  daily observations to
avoid additional complications induced by averaging.     
}
 over the week, since
  the producers try to maximise the accumulated income and buyers are
  likewise minimising their accumulated expenses. 
Had we instead considered, say, the hourly price at hour 24 each
Sunday, our results would not necessarily say much about price
information flow between the weekly price levels.

\begin{figure}[htb]
  \centering
  \includegraphics[width=\linewidth]{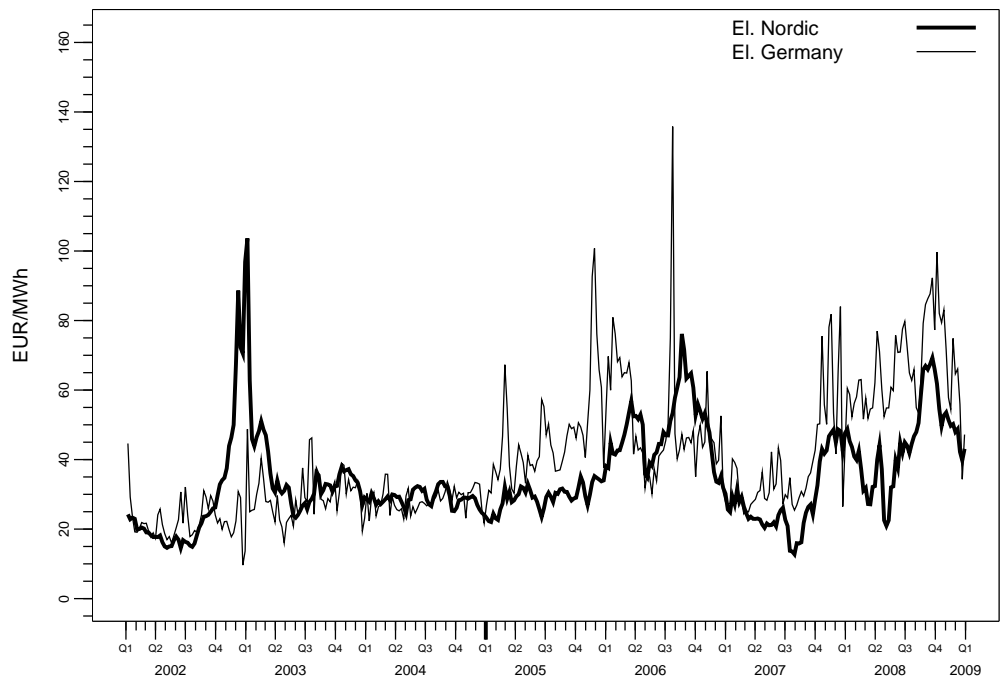}
  \caption{Weekly electricity spot prices, Nordic  and German.}
  \label{fig:data:el}
\end{figure}

\begin{figure}[htb]
  \centering
  \includegraphics[width=\linewidth]{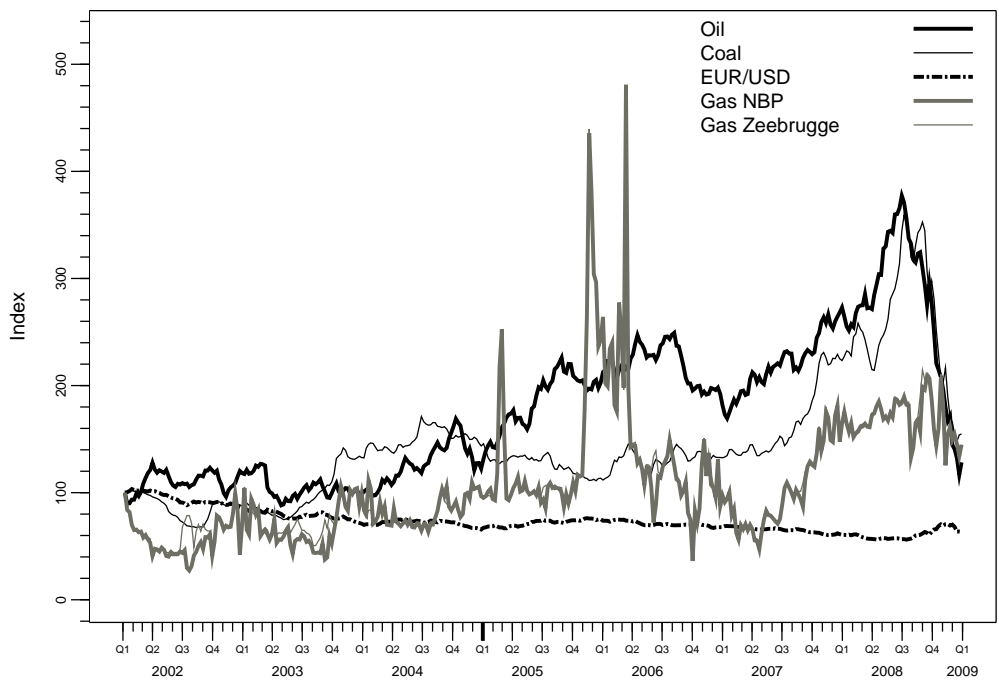}
  \caption{Weekly oil, coal and gas prices, and the EUR/USD exchange rate. All price series are given in Euro and indexed to 100 at Week 1 January 2002.}
  \label{fig:data:com}
\end{figure}

\begin{figure}[htb]
  \centering
\subfigure[Water reservoir levels, Nordic (Norway+Sweden).]{
  \includegraphics[width=0.7\linewidth]{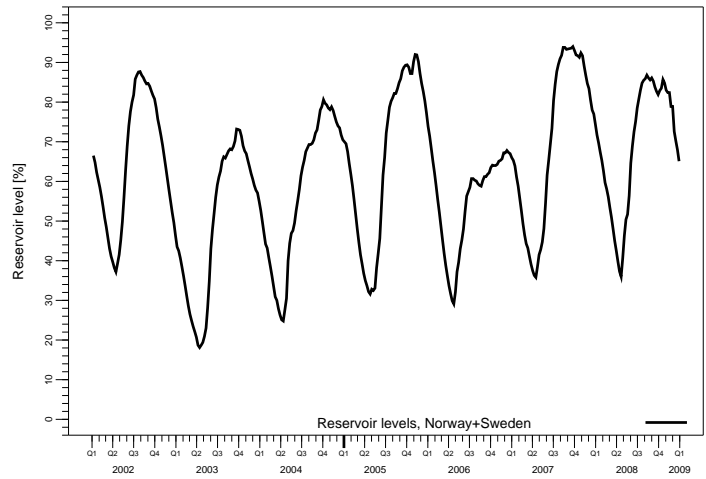}}
\subfigure[Electricity production from wind plants in Germany.]{
  \includegraphics[width=0.7\linewidth]{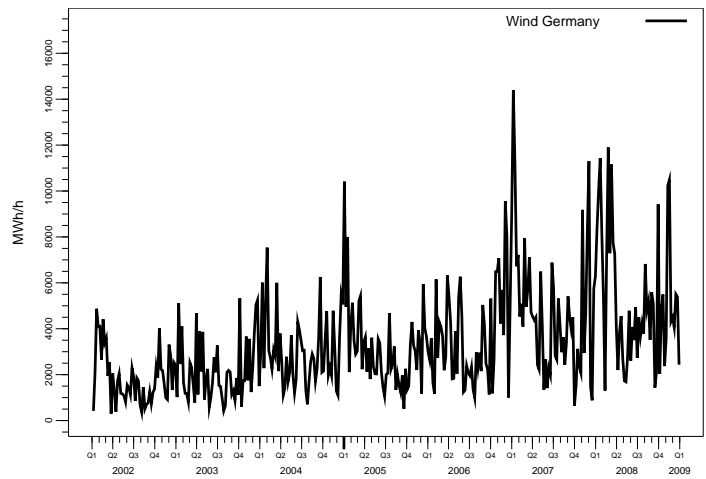}}
  \caption{Weekly values of exogenous variables. 
}
  \label{fig:data:w}
\end{figure}

An overview of the data is given in Table \ref{tab:data}, and
they are displayed in Figure \ref{fig:data:el}--\ref{fig:data:w}.

We have included two gas prices here: Zeebr\"ugge and NBP,
representing continental Europe and the United Kingdom,
respectively. A few other gas markets are more relevant for the German
electricity market than the Zeebr\"ugge gas market, but the historic
data period is then not long enough. The Zeebr\"ugge and NBP gas
markets are connected through the Bacton--Zeebr\"ugge interconnector
\citep{micola07}. We expect the Zeebr\"ugge gas market to be more
important than the NBP market for the electricity price formation, since it is closer to the German (and Nordic) market. 

We will treat German wind and Nordic reservoir levels as exogenous
variables in  \eqref{eq:vecm}, which to some extent is debatable. The
electricity market does not influence the wind itself, but may have
contributed to the long term increase in wind power mills. Similarly,
the electricity market does not influence the inflow into water
reservoirs, but the water reservoirs are ideally used when prices are
high. Still, we find it most correct to treat these two variables as
exogenous.

All time series, except reservoir levels,  were
log-transformed. Since the reservoir levels are bounded by 0 and
100\%, they were logit-transformed. Next, all transformed variables except the oil
price, were seasonally adjusted by subtracting a seasonal term
\begin{align*}
\lambda_t = \beta^{(0)} + \sum_{j=1}^2{\beta^{(1)}_j \sin \left( \frac{2 \pi
j t}{52} \right)
+ \beta^{(2)}_j \cos \left( \frac{2 \pi j t}{52} \right), }
\end{align*}
which was estimated by least squares regression.
This was done in order to have variables that represent deviations
from a normal level.

\begin{table}[ht]
\begin{center}
\begin{tabular}{r|l|l}
  Data & Description & Resolution\\
  \hline
  Nordic el.\ price & Nord Pool system & Weekly, average spot price \\
  German el.\ price & European Energy Exchange (EEX) & Weekly, average spot price \\
  Oil price & Brent crude, International& Weekly, average spot price\\
&  Petroleum Exchange (IPE)  &  \\
  Gas price 1 & National balancing point   & Weekly, average spot price \\
& (NBP), UK & \\
  Gas price 2 & Zeebr\"ugge,  Belgium  & Weekly, average spot price \\ 
  Coal price & CIF ARA, Northwest Europe  & Weekly, average physical price \\
  EUR/USD & Exchange rate & Weekly, average rate\\
  Water  Nordic &  Reservoir levels, Norway+Sweden & Values for each Monday\\
Wind Germany & Electricity production, wind plants & Weekly, average production \\
  \end{tabular}
\caption{Data overview. The data range from the first week of 2002 to the
  last week of 2008, in total 365 weekly values for each of the variables. Since German wind
  production is not publicly available, the German wind
  production data were calculated by  Point Carbon (http://www.pointcarbon.com).} 
\label{tab:data}
\end{center}
\end{table}

\clearpage

\section{Results}
\label{sec:results}

\subsection{Time series analysis}

Using the criteria of Akaike, Schwarz and Hannan and Quinn
\citep{claeskens08}, the optimal lag order of the unrestricted VAR
with the two exogenous variables \eqref{eq:var} was
found to be two (Table \ref{tab:varselect}), even though Akaike's
criterion was almost as good for three lags.  
 Phillip-Perron unit root tests indicate that there is no evidence
for stationarity of the series Oil, EURUSD and Coal. The same tests performed on the first differences
indicate that these are stationary. Table \ref{tab:pp-test} shows the
test statistics and the p-values for each of the series.  
\begin{table}[ht]
\centering
\begin{tabular}{|r|rrrrr|}
\hline
 & \multicolumn{5}{|c|}{Lag order}\\ 
Model selection criterion & 1 & 2 & 3 & 4 & 5\\ \hline 
Akaike & -38.96 & -39.55 & -39.54 & -39.51 & -39.38\\ 
Hannan and Quinn & -38.65 & -39.03 & -38.81 & -38.57 & -38.23\\ 
Schwarz & -38.19 & -38.25 & -37.71 & -37.14 & -36.48\\ 
\hline
\end{tabular}
\caption{Akaike, Schwarz and Hannan and Quinn model selection criteria for the optimal lag order of the VAR with a constant.}
\label{tab:varselect}
\end{table}
\begin{table}[ht]
\centering
\begin{tabular}{|r|rr|rr|}
\hline
 & \multicolumn{2}{|c|}{Original data} & \multicolumn{2}{|c|}{First differences}\\ 
 & statistic & p-value & statistic & p-value\\ \hline 
ElNordic &  -3.6685 &    0.0050  & -14.5090  &    0.0000\\ 
ElGermany & -4.5652 &    0.0002  & -26.8163  &    0.0000\\ 
Oil &       -1.5193 &    0.5229  & -15.2474  &    0.0000\\ 
EURUSD &   -2.1031 &    0.2437  & -14.9070  &    0.0000\\ 
Coal &      -1.3370  &   0.6132  & -10.8962  &    0.0000\\ 
GasNBP &   -3.1755  &   0.0223  & -23.8243  &    0.0000\\ 
GasZEE &   -3.0617  &   0.0304  & -23.2813  &    0.0000\\ 
\hline
\end{tabular}
\caption{Stationarity test using the Phillip-Perron unit root test: Test statistics for both the original time series (on log scale) and the first differences of the time series (on log scale), when testing for stationarity. The null hypothesis is that the series has a unit root, i.e.~that they are non-stationary.}
\label{tab:pp-test}
\end{table}

We have also used another test of stationarity, the KPSS test \citep{KPSS1992}, where the null hypothesis is that each of the series is stationary. All time series, except for ElNordic, are rejected at a $1\%$ significance level. ElNordic is rejected at a $5\%$ significance level. The KPSS test on first differences indicate that all time series are level-stationary when differenced. 

We use the trace test to determine the number of cointegrating
vectors, and the Schwarz criterion for determining whether the
constant is within or outside the cointegration space. The trace test
indicates that there are three cointegrated vectors at a $1\%$
significance level,  and the Schwarz
criterion indicates that the constant is inside the cointegrating
space\footnote{The Schwarz criterion indicates that there are four
  cointegrating vectors, but we proceed with the trace test's
  conclusion.}. Table \ref{tab:coint-const} shows the test statistics
and the critical values for different ranks when the constant is
within the cointegration space. Table \ref{tab:bic-constant-within}
shows the Schwarz criterion for the constant within and outside the
cointegration space, when the cointegration rank is three. 
 Since the  cointegration tests can be  sensitive to the lag
  structure of the VAR \citep{kasa92}, we have repeated the trace test
  for  up to five lags in the VAR. In any case, and regardless of
  whether the constant is within or outside the cointegration space,
  we conclude that the number of cointegrating vectors is three.
\begin{table}[ht]
\centering
\caption{Trace test of cointegration, when the constant is within the cointegration space. }
\begin{tabular}{|c|r|rrr|}
\hline
Rank & Trace test & \multicolumn{3}{|c|}{Critical values}\\ 
(r) & statistic & $10\%$ & 5\% & 1\%\\ \hline 
$r \leq 6$ &   3.72 &   7.52 &   9.24 &  12.97\\ 
$r \leq 5$ &  13.92 &  17.85 &  19.96 &  24.60\\ 
$r \leq 4$ &  29.39 &  32.00 &  34.91 &  41.07\\ 
$r \leq 3$ &  59.77 &  49.65 &  53.12 &  60.16\\ 
$r \leq 2$ & 102.23 &  71.86 &  76.07 &  84.45\\ 
$r \leq 1$ & 172.97 &  97.18 & 102.14 & 111.01\\ 
$r = 0$ & 280.18 & 126.58 & 131.70 & 143.09\\ 
\hline
\end{tabular}
\end{table}

\begin{table}[ht]
\centering
\begin{tabular}{|r|r|}
\hline
 & Schwarz loss\\ \hline 
Constant within & -3487.45\\ 
Constant outside & -3482.29\\ 
\hline
\end{tabular}
\caption{The Schwarz loss when the constant is inside and outside the cointegration space, for the case of three cointegrated vectors.}
\label{tab:bic-constant-within}
\end{table}

A cointegrating vector is a stationary linear combination of possibly non-stationary vector time-series components. This combination might consist of only one of the series, which then must be stationary. It is interesting to test if this is the case, especially since the Phillip-Perron test suggests that several of the series are stationary. Table \ref{tab:itself} shows the p-values of this test applied to each of the series, whose null hypothesis is that the series is by itself one of the cointegrating vectors. We see that all series are rejected, indicating that none of the cointegrating vectors consist of only one of the series.

\begin{table}[ht]
\centering
\begin{tabular}{|r|r|}
\hline
 & P-value\\ \hline 
ElNordic & 0.0005\\ 
ElGermany & 0.0000\\ 
Oil & 0.0000\\ 
EURUSD & 0.0000\\ 
Coal & 0.0000\\ 
GasNBP & 0.0002\\ 
GasZEE & 0.0001\\ 
\hline
\end{tabular}
\caption{Test of whether a series by itself is one of the cointegrating vectors.}
\label{tab:itself}
\end{table}

After fitting a VEC model with rank three to our data, see
\eqref{eq:vecm}, we perform a normality test on the residuals. We use
the Jarque-Bera test which tests for normality in both the univariate
and multivariate case. The test rejects the null hypothesis of
normality for each univariate series and for the multivariate case as
well. In addition, an investigation of the residuals showed no
significant auto-correlation between the residuals.
Thus, the assumption of independent and non-Gaussian residuals is not unreasonable, and LiNGAM can be used.

A weak exogeneity test is performed, which tests the null hypothesis that each of the series does not respond to disturbances or shocks in the cointegration space, i.e.~that the series is unresponsive to the deviations from the long-run relationships. This test is performed on $\boldsymbol{\alpha}$, more specifically, for one particular series, we test whether the corresponding row in $\boldsymbol{\alpha}$ (and hence in $\boldsymbol{\Pi}$) is zero. 

Further, an exclusion test is performed, which tests the null hypothesis that a particular series is not in the cointegration space. This test is performed on $\boldsymbol{\beta}$, also here testing for a zero row. For more details on tests on $\boldsymbol{\alpha}$ and $\boldsymbol{\beta}$, see e.g.~\cite{Juselius2006}.

Table \ref{tab:exo-exc} shows the p-values for both the weak exogeneity test and the exclusion test. We see that ElNordic, ElGermany and GasNBP are rejected at a $3\%$ or lower significance level in the weak exogeneity test, meaning that the long-run relationships in the data are important for these series, whereas for the other series there is not evidence for this. In the exclusion test, ElNordic, ElGermany, EURUSD, GasNBP and GasZEE are rejected. Hence, there is strong evidence that these series are included in the long-run relationships. An exclusion test is also performed on the constant term, which results in a rejection of the null hypothesis at a $2\%$ significance level. This agrees with the Schwarz conclusion in Table \ref{tab:bic-constant-within}.


\begin{table}[ht]
\centering
\begin{tabular}{|r|r|r|}
\hline
 & Weak exogeneity & Exclusion\\ \hline 
ElNordic & 0.0065 & 0.0001\\ 
ElGermany & 0.0000 & 0.0000\\ 
Oil & 0.3722 & 0.1458\\ 
EURUSD & 0.8023 & 0.0019\\ 
Coal & 0.5012 & 0.1388\\ 
GasNBP & 0.0298 & 0.0000\\ 
GasZEE & 0.1866 & 0.0000\\ 
\hline
\end{tabular}
\caption{The p-values for the weak exogeneity test and the exclusion test, whose null hypothesis is that 
a particular series does not respond to shocks in the cointegration space, and that a particular series 
is not in the cointegration space, respectively.}
\label{tab:exo-exc}
\end{table}

\begin{figure}[h!]
  \centering
  \includegraphics[width=\linewidth]{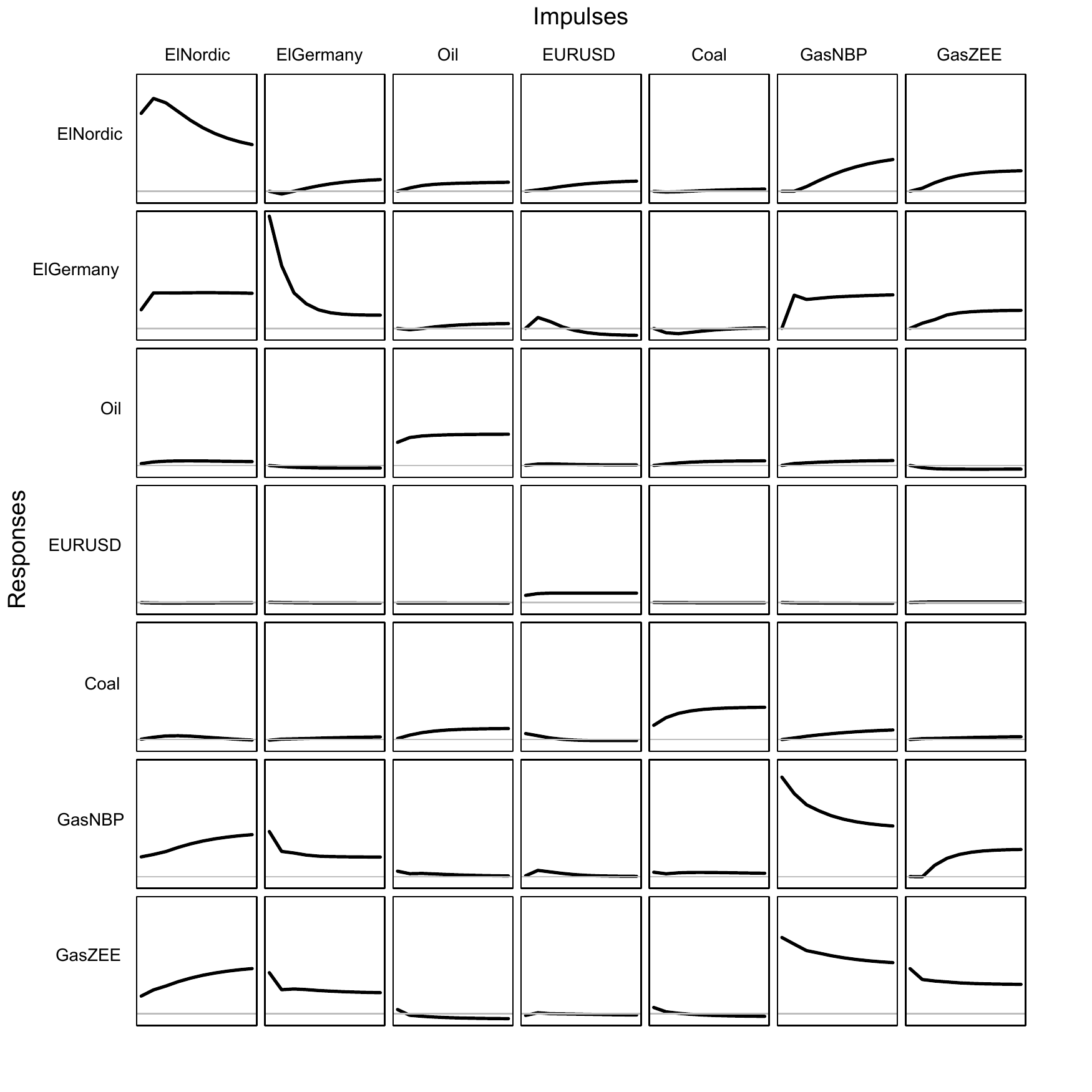}
  \caption{Impulse response plot: Each column shows the up to ten week responses in all series to a one-time-only shock in the series listed in the column header.}
  \label{fig:irf-plott}
\end{figure}

Figure \ref{fig:irf-plott} displays the impulse responses for all series, i.e.~the responses of each series to a shock in each series. Each column shows the up to ten week responses of all series caused by an impulse (a one-time-only shock) in one of the series (the column headers show the impulses, whereas the row headers show the responses). The responses are normalised so that they can be compared with each other. 

In order to get the impulse responses, the causal ordering
  among variables is needed. For Figure \ref{fig:irf-plott}, we have
  followed the standard approach and used Bernanke ordering
  \citep{bernanke86}. The innovations are written as a function of
  more fundamental, internally orthogonal sources of variation,
  $\boldsymbol\nu_t$, given by $\e_t=
  \widetilde{\boldsymbol A}^{-1}\boldsymbol\nu_t$, where $\widetilde{\boldsymbol A}$ is a matrix representing how the innovations $\e_t$ are caused by orthogonal variation in each variable. Alternatively, we could here have used our LiNGAM based ordering.

As seen on the diagonal, all series respond positively to their own shocks, and except for EURUSD, these responses are also strong. ElGermany responds quickly and strongly to shocks in ElNordic, whereas there is not much impulse response the other way around. GasNBP and GasZEE have a slowly increasing response to a shock in ElNordic, whereas they respond much quicker to a shock in ElGermany.

ElNordic is mostly affected by impulses in GasNBP and GasZEE, and also here, these responses are slowly increasing over time.
Besides being affected by impulses in ElNordic, ElGermany is also affected by impulses in GasNBP and GasZEE. The response caused by a shock in GasNBP is much quicker for ElGermany than for ElNordic. Further, GasZEE is more affected by a shock in GasNBP than the other way around. 

Finally, we see that Oil, EURUSD and Coal are neither causing any significant responses in the other series, nor responding to shocks in any of the other series.  

\subsection{Learning contemporaneous and time-lagged causal DAGs}

 In the following we investigate the instantaneous causal ($\B_0$) and
 lagged ($\B_1$ and $\B_2$) effects.

The time series were standardised before the following DAG analysis, enabling direct comparison of the strengths of causal effects. 
$\B_0$, $\B_1$ and $\B_2$ were then estimated as described in Section~\ref{sec:instlag}. 
Insignificant edges of $\B_0$ were removed using the resampling method described in Section 6.3 of~\cite{shimizu2006lng}. 
As seen in \eqref{eq:B_lagged} in Section~\ref{sec:instlag}, the time-lagged effects $\B_\tau, \tau=1,2$, 
depend on both $\B_0$ and the matrix $\M_\tau$ of pure autoregressive effects. Therefore, the resampling 
method used for $\B_0$ is not available for the time-lagged effects, and it is not clear from Equation~\eqref{eq:B_lagged} how we could assess significance e.g.~using p-values. However, since the data are standardised, we may simply use a cutoff in effect 
size as our significance threshold. 
 We have chosen to remove all effects from  $\B_1$ and $\B_2$ that are
 smaller in absolute value than the $70\%$ absolute value quantile of all the elements in
 $\B_1$. 

\begin{figure}[h!]
  \centering
  \includegraphics[width=8cm]{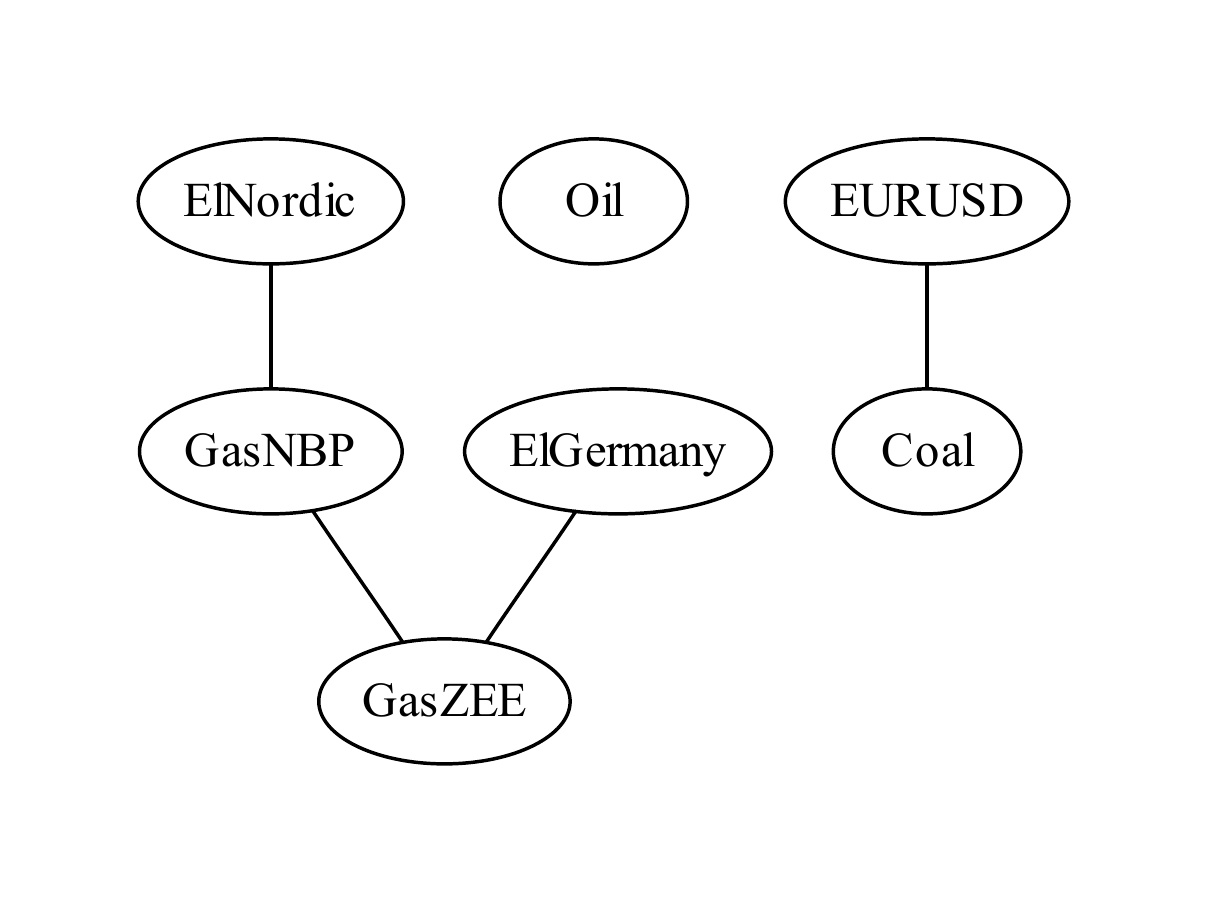}
  \caption{The instantaneous causal effects obtained using the GES algorithm in \cite{tetradManual}.}
  \label{fig:ges}
\end{figure}

To illustrate the advantages of the use of the LiNGAM methodology, we
show instantaneous effects estimated using the GES algorithm, as
implemented in \cite{tetradManual}.
 The results are shown in the partially directed acyclic graph
(PDAG) in Figure~\ref{fig:ges}. The PDAG shows the entire equivalence
class as a single graph. Having a directed edge in the PDAG means that
this edge has the same orientation for all DAGs in the equivalence
class. Undirected edges in the PDAG have different orientations for
different members of the Markov equivalence class. Note that the PDAG
in Figure~\ref{fig:ges} is completely undirected, so no directions of
causal influences can be determined in this case.  Figure~\ref{fig:ges} shows an association between 
 the  coal price and the EURUSD. No price information seems to flow
 to or from the oil price, while the Nordic and German electricity prices seem
 to be connected through the two gas prices.

\begin{figure}[h!]
  \centering
  \includegraphics[width=0.8\linewidth]{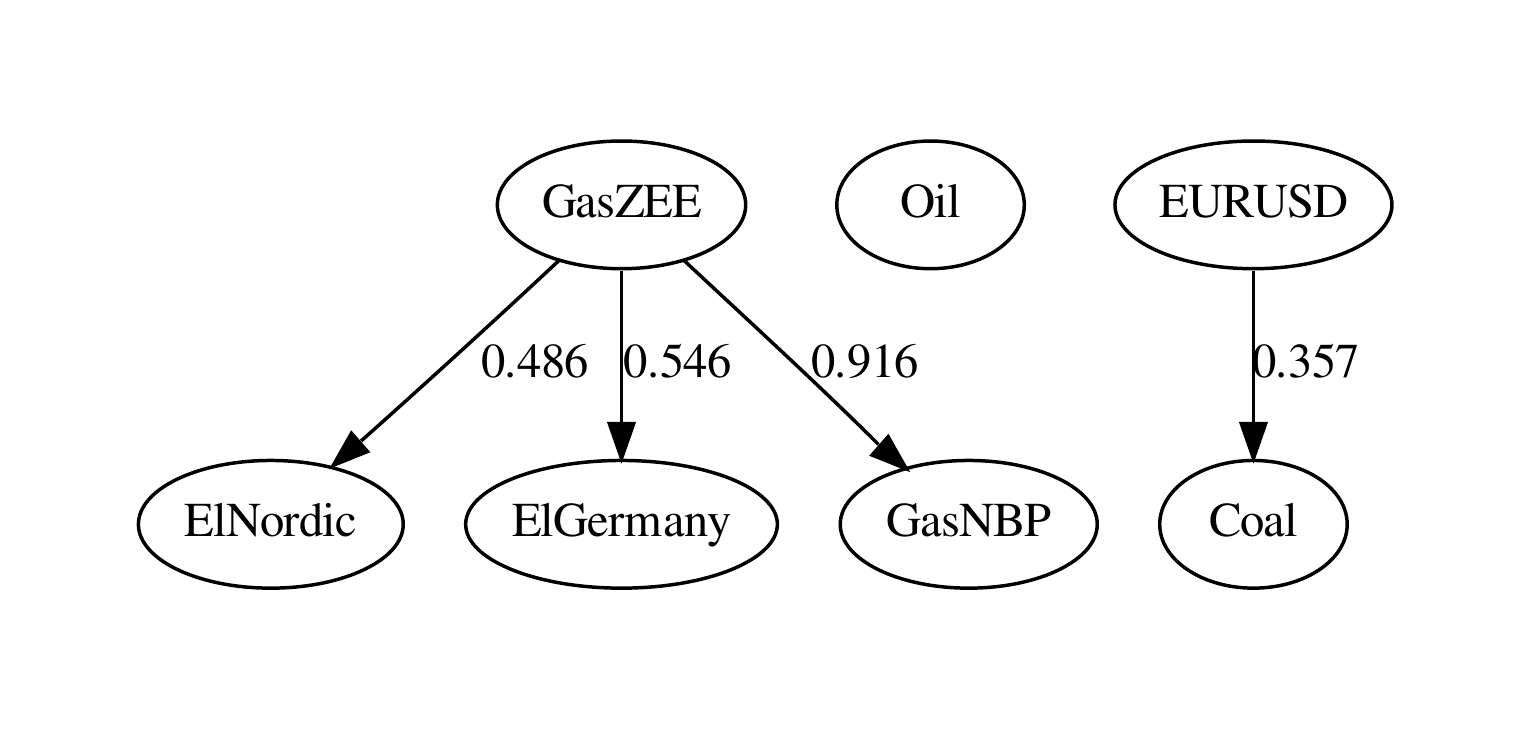}
  \caption{$\B_0$: The instantaneous causal effects.}
  \label{fig:B0}
\end{figure}

Figure \ref{fig:B0} shows the graphical representation of $\B_0$,
estimated using LiNGAM, as described in Section~\ref{sec:instlag}. Most
of these instantaneous effects are intuitively reasonable. The main difference
between the DAG $\B_0$ and the PDAG obtained using the GES algorithm
in Figure~\ref{fig:ges} is that the latter lacks directions of the
edges. Again, information does not flow to/from the oil price. As
expected, the arrow goes from EURUSD to coal prices. Information flows
from GasZEE to GasNBP, ElNordic and ElGermany. This is partly
surprising, but we should keep in mind that the Nordic reservoir
levels and German wind have already been accounted for in the model,
and it might be that, contemporaneously, GasZEE plays an important role.
 
\begin{figure}[h!]
  \centering
  \includegraphics[width=0.8\linewidth]{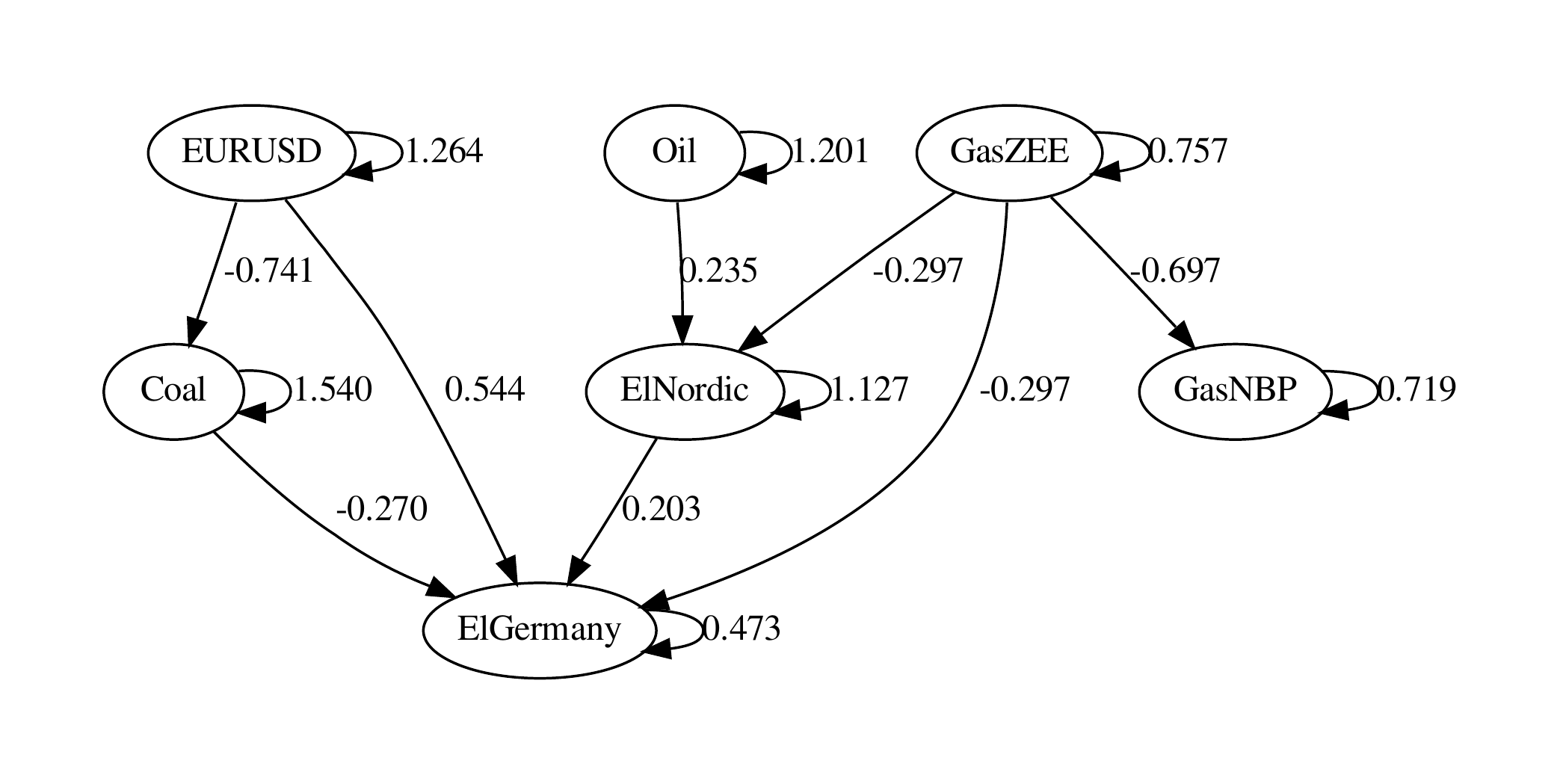}
  \caption{$\B_1$: The causal effects with lag one. The smallest
effects have been removed.}
  \label{fig:B1}
\end{figure}

\begin{figure}[h!]
  \centering
  \includegraphics[width=0.8\linewidth]{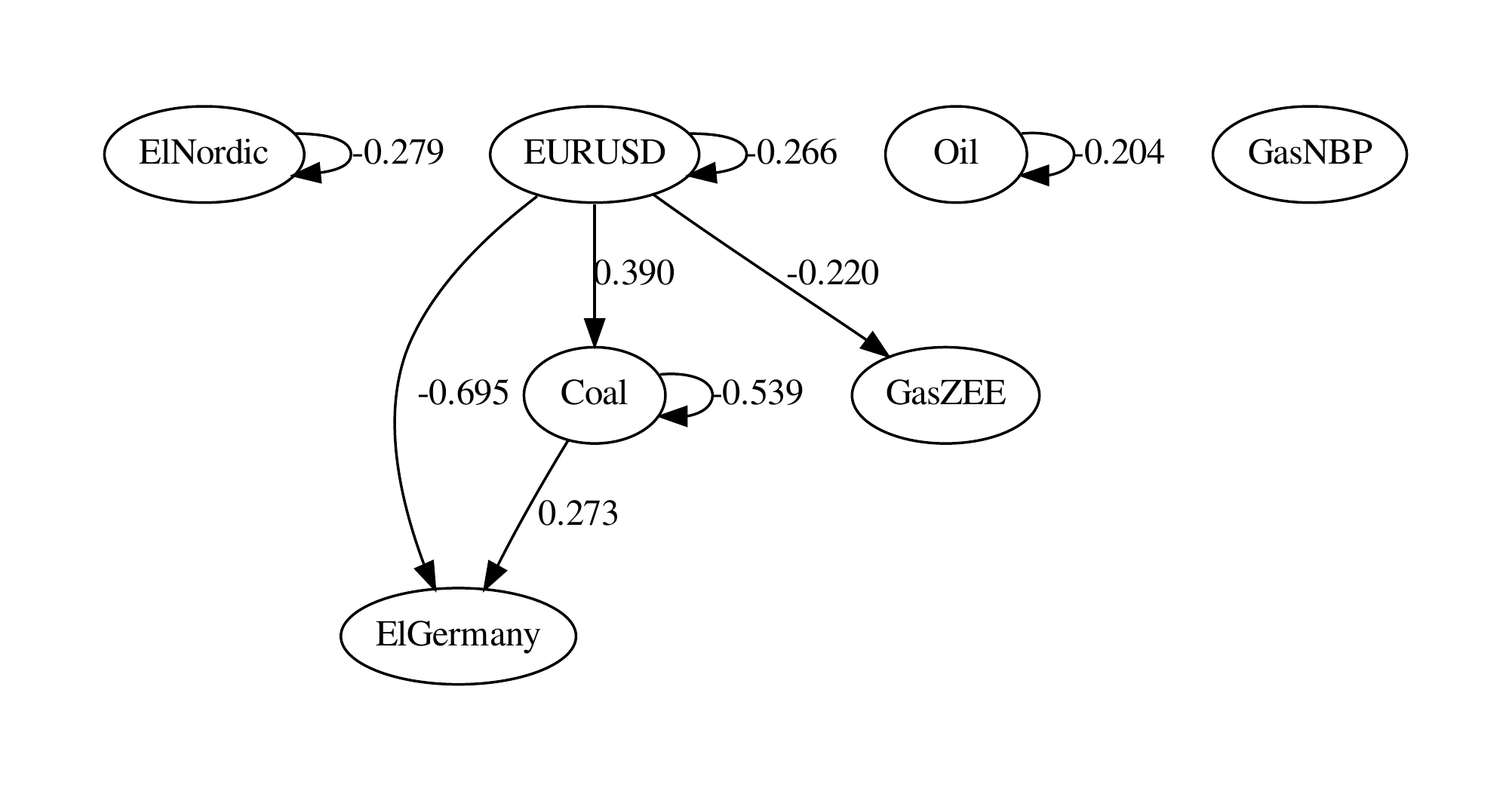}
  \caption{$\B_2$: The causal effects with lag two. The smallest
effects have been removed.}
  \label{fig:B2}
\end{figure}

Figure \ref{fig:B1} and \ref{fig:B2} show the graphical representations
of $\B_1$ and $\B_2$, respectively.
Note that these graphs are directed, but cyclic, so they are not
DAGs. 
This is natural for time-lagged relationships.
We see that all
variables influence themselves at time lag one, and that ElNordic, EURUSD, Coal and Oil even influence themselves at time lag two.    

At lag one ($\B_1$), ElGermany is (mainly) influenced
directly and indirectly by EURUSD and GasZEE, indirectly by Oil, and
directly by Coal and ElNordic. GasNBP is influenced by GasZEE (and
GasNBP itself), but influences nothing else. Note, however, that some
of the effects are quite small, except for the EURUSD $\rightarrow$ Coal, EURUSD $\rightarrow$ ElGermany and GasZEE $\rightarrow$ GasNBP relationships. At lag two ($\B_2$), there
are fewer strong effects, except that EURUSD seems to play an important
role. 

\section{Discussion}
\label{sec:disc}

Using time series models combined with new advances in causal
inference, we have studied how  dynamic price information flows among Northern European electricity  spot prices and prices of major electricity  generation fuel sources. 
Applying our methods to weekly 
Nordic and German electricity prices, and oil, gas and coal prices, as
well as German wind power and Nordic water reservoir levels, we have estimated a causal
model for the price dynamics, both for contemporaneous and lagged
relationships. 

We find that the oil price, coal price and EUR/USD exchange rate are
non-stationary, while Nordic and German electricity prices, as well as
British and Zeebr\"ugge gas prices are stationary. 
Our results can be compared with the results from
\cite{mjelde2009mia}, who study the US market, even though we have
treated Nordic water reservoir levels and German wind power as
exogenous variables.
There are a few
noteworthy similarities and differences. \citeauthor{mjelde2009mia}
include both peak, and off-peak prices, while we consider base
prices. Note, however, that the peak/off-peak difference in the Nordic
electricity market is less pronounced due to the very flexible hydro power.
Contrary to  \citeauthor{mjelde2009mia}, we find only positive
innovation shock responses, for example from natural gas to coal,
where there is a negative response in the US study. We both find a
strong connection between gas and electricity prices. In contemporaneous
time, we find a causal link from (Zeebr\"ugge) gas prices to the electricity
markets, while the US study gives the opposite conclusion. 
We find that coal and EURUSD together stand alone in contemporaneous time. In the US study, where the exchange rate is not included in the analysis (since all prices are in USD), coal stands alone in contemporaneous time. 
We find that even oil stands alone in contemporaneous time, which could be explained
by the difference in European and US gas markets \citep{hobakhaff08},
even though they may converge due to the increase in liquefied natural
gas trade  \citep{neumann09}. 
As with \cite{mjelde2009mia}, we find that all price series are
cointegrated with a few cointegrating vectors (three in our case). At
longer horizons, electricity prices and British gas prices adjust themselves to establish the
equilibrium price level, since oil, coal, continental gas and EUR/USD are found to be weakly
exogenous. In our analysis, however, and contrary to the US study, the exclusion test casts some doubt on whether the oil and coal prices are part of the cointegrating space.

Generally, the British gas prices are not
important for the electricity markets when the Zeebr\"ugge gas price is
included, which is expected, since the Zeebr\"ugge gas market is closer to the
electricity generation and grid.
The fact that coal prices do not play an important role in
 contemporaneous time in our analysis, while gas does, could first of
 all be because we have employed the more liquid CIF ARA price, while
 local producers may pay a different price, which may also partly be
 the case for the Zeebr\"ugge gas prices. Second, the coal price has
 a low volatility compared to the gas and electricity prices, and
 naturally reacts more slowly to peak demand, since the coal prices' influence
 is affected by transportation time and costs. Third, there has been
 speculation that the oil and gas markets in Europe are decoupling
 (see e.g.\ \cite{OilGasUK}), which could also partly explain why the
 oil and gas prices play different roles in this Northern European
 commodity price game.


In our view, there are two main methodological advantages of our approach, as
compared to  previous work~\citep{mjelde2009mia,park2006pda,park2008pia}. First, we are able to identify one unique contemporaneous graph, as opposed to a Markov
  equivalence class (which might be large).
Second, we are able to properly and coherently deal with both
  instantaneous and time-lagged effects in the same
  analysis. \citet{park2006pda} (p.~97) state that ``in contrast to the
  directed graph analysis, forecast error variance decomposition and
  impulse response functions allow for analysis of dynamic information
  flows over time'', i.e.~in their view, DAGs are only applicable for
  analysing instantaneous effects. We have shown that DAGs are in fact
  useful for combining time-lagged and instantaneous effects.

Implicit in our premise of statistically independent
errors/residuals is the assumption of having no unobserved
confounders: Any unmeasured common cause of any two of our variables
would skew our results and create a dependence. It is possible to include latent variables in the LiNGAM model~\citep{hoyer08hidden},
but we have seen this as out of the scope of our paper, due to the added complications of dealing with time series data. 

Our approach is a first attempt at a causal model for the price
dynamics, and can be improved in many ways. Future work could include
non-linear causal discovery~\citep{hoyer09},  incorporating
possible effects of stochastic volatility and investigating the price
dynamics on a finer time scale, for example with daily instead of
weekly price series.

\subsection*{Acknowledgements}
This work is funded by Statistics for Innovation, (sfi)$^2$, one of the 14 Norwegian Centres for Research-based Innovation.
 We thank
Norsk Hydro for supplying the data, and in particular R{\o}nnaug S{\ae}grov
Mysterud for useful discussions. We are grateful to Arnoldo Frigessi for helpful comments. We appreciate Patrik O.~Hoyer's help
with the LiNGAM methodology and software.






\newpage
\bibliographystyle{elsarticle-harv}  
\bibliography{el}     

\end{document}